\documentclass[12pt,amsfonts]{article}   
\usepackage{amsfonts}
\def\t{\textstyle}        
\def\one{1\hskip-.37em 1}                 

\def\tp{{\widetilde p}}
\def\tq{{\widetilde q}}
\def\tb{\tilde{\beta}}
\def\half{{\textstyle{\frac{1}{2}}}}

\def\b{\beta}

\def\H{{\cal H}}

\def\p{\phi}
\def\th{\theta}
\def\H{{\cal H}}

\def\ra{\rightarrow}

\def\tint{{\textstyle\int}}

\def\s{\hskip.08em}
\def\d{\partial}

\def\b{\begin{eqnarray*}}  
\def\e{\end{eqnarray*}}    
\def\bn{\begin{eqnarray}}  
\def\en{\end{eqnarray}}   
\def\<{\langle}
\def\>{\rangle}

\def\no{\nonumber}

\def\{{\lbrace}
\def\}{\rbrace}

\title{Enhanced Quantization: A Primer}
\author{John R. Klauder\footnote{Email: klauder@phys.ufl.edu}\\
Department of Physics and\\Department of Mathematics\\
University of Florida\\
Gainesville, FL 32611-8440}
\date{ }

\begin{document}
\maketitle
\begin{abstract}
Although classical mechanics and quantum mechanics are separate disciplines, we live
in a world where Planck's constant $\hbar>0$, meaning that the classical and quantum world views must actually {\it coexist}.
Traditionally, canonical quantization procedures postulate a direct linking of various $c$-number and $q$-number
quantities that lie in disjoint realms, along with the quite distinct interpretations given to each realm.
In this paper we propose a different association of classical and quantum quantities that renders classical theory
a natural subset of quantum theory letting them coexist as required. This proposal also shines light on alternative linking assignments of classical and quantum quantities that offer different perspectives on the very meaning of quantization. In this paper we focus on elaborating the general principles, while elsewhere we have published several examples of what this alternative viewpoint can achieve; these examples include removal of singularities in classical solutions to certain models,
and an alternative quantization of several field theory models that are trivial when quantized by traditional methods but become well defined and nontrivial when viewed from the new viewpoint.
\end{abstract}
\subsection*{Introduction}
Given a quantized system, there are various ways to describe it: Schro\"odinger's equation; Heisenberg's
formulation; Feynman's path integral; etc., all of which lead to the same results. However, scant attention is paid any longer to just how a classical theory is connected to its quantum theory, i.e., the very process of quantization itself. How a classical and quantum system are related is the principal topic of the present work, and our procedures differ from the standard approach. In so doing,
we are able to cement some older suggestions into a concrete prescription. In the author's earlier works \cite{CRT2,CRT3,KLASK}, for example,  similar ideas were advanced but they lacked a consistent, unified approach, a deficiency finally rectified in this paper. One additional property that helps crystalize this consistency is an insistence on basic kinematical variables that are self adjoint and not merely Hermitian; the meaning of and the reasons behind this choice are given below. The procedures developed here have already been applied to certain systems, and some of the most interesting ones are field theories for which traditional quantization methods yield triviality while alternative procedures modeled on those within this paper yield nontrivial results. References for several previous applications appear later in the paper.

We call the alternative approach to quantization presented in this paper: {\it enhanced quantization}.

\subsection*{Enhanced Canonical Quantization}
We use the language of one-dimensional classical phase-space mechanics with well-chosen canonical coordinates $p$ and $q$, which, of course, fulfill the Poisson bracket  $\{q,p\}=1$. Likewise, we adopt the language of abstract quantum mechanics that includes basic Hermitian operators such as $P$ and $Q$, for which
 $[Q,P]=i\s\hbar\s\one$, and Hilbert space
vectors $|\psi\>$, their adjoints $\<\psi|$, and their inner products such as $\<\p|\psi\>$, which is linear in the right-hand vector. The kinematical part of canonical quantization  ``promotes'' $q\ra Q$ and $p\ra P$, for which
$(a\s q+b\s p)\ra (a\s Q+b\s P)$,
where $a$ and $b$ are fixed coefficients. This form of assignment extends to functions of the basic operators such as $H_c(p,q)\ra \H=H_c(P,Q)$, at least to leading order in $\hbar$ in the right coordinates. If we interpret $H_c(p,q)$ as the classical Hamiltonian, it follows that the classical dynamical equations are given by $dq/dt=\d H_c(p,q)/\d p$ and
$dp/dt=-\d H_c(p,q)/\d q$, with solutions determined by initial conditions $(p_0,q_0)$ at $t=0$. In turn, dynamics in the quantum theory may be expressed in the form $i\hbar\s\s\d \s|\psi\>/\d \s t=\H\s|\psi\>$, with a solution fixed by
the initial condition $|\psi_0\>$ at $t=0$. The two sets of dynamical equations of motion arise from separate classical (C) and quantum (Q) action functionals given (with ${\dot q}\equiv dq/dt$, etc.), respectively, by
  \bn A_C\hskip-1.4em&& =\tint_0^T[\s p(t)\s\s {\dot q(t)}-H_c(p(t),q(t))\s]\,dt\;,\no\\
    A_Q\hskip-1.4em&&=\tint_0^T\<\psi(t)|\s[\s i\s\hbar\s \d/\d t-\H(P,Q)\s]\s|\psi(t)\>\,dt\;, \label{AQ} \en
    where we may assume that the vectors $|\psi(t)\>$ are normalized to unity. The formalisms of Heisenberg and Feynman are equivalent to the foregoing, and they also deal with distinct realms for classical and quantum mechanics. We now ask: Is there another way to quantize in which the classical and quantum realms coexist?

    We initially chose the basic quantum kinematical variables $P$ and $Q$ that were assumed to be
    {\it Hermitian} [e.g., $P^\dag=P$ on the domain ${\cal D}(P)\subseteq {\cal D}(P^\dag)$], but now we confine attention to those operators that are {\it self adjoint} [e.g., $P^\dag=P$ on the domain ${\cal D}(P)= {\cal D}(P^\dag)$]. This restriction is introduced because only self-adjoint operators can serve as generators of unitary one-parameter groups. Thus, we can introduce two, basic unitary groups defined by
    $U(q)\equiv \exp(-i\s q\s P/\hbar)$ and $V(p)\equiv \exp(i\s p\s Q/\hbar)$ that satisfy
         \bn U(q)\s V(p)= e^{\t -i\s p\s\s q/\hbar}\,V(p)\s U(q)\;, \en
         which guarantees a representation of $P$ and $Q$, with a spectrum for each operator covering the whole real line, unitarily equivalent to a Schr\"odinger representation, provided the operators are irreducible. We choose
         a fiducial unit vector $|0\>$ that satisfies the relation $(Q+i\s P)\s|0\>=0$, which implies that $\<0|\s P\s|0\>=\<0|\s Q\s|0\>=0$, and use it to define  a set of canonical coherent states
          \bn |p,q\>\equiv e^{\t -iq\s P/\hbar}\,e^{\t ip\s Q/\hbar}\,|0\>\;,\label{ccs}\en
          for all $(p,q)\in{\mathbb R}^2$, which are jointly continuous in $p$ and $q$; for convenience, we let $P$ and $Q$ (and thus $p$ and $q$) have the same dimensions, namely, those of $\hbar^{1/2}$. It may be helpful to picture the coherent states as a continuous, two-dimensional sheet of unit vectors coursing through an infinite-dimensional space of unit Hilbert space vectors.

          With regard to Eq.~(\ref{AQ}), Schr\"odinger's equation results from stationary variation of $A_Q$ over arbitrary histories of Hilbert space unit vectors $\{|\psi(t)\>\}_0^T$, modulo fixed end points. However, {\it macroscopic} observers studying a {\it microscopic} system cannot vary vector histories over such a wide range. Instead, they are confined to moving the system to a new position $(q)$ or changing its velocity $(p)$ [based on ${\dot q}=\d H(p,q)/\d p\s$]. Thus a macroscopic observer---or better a {\it classical observer}---is restricted to vary only the quantum states $|p(t),q(t)\>$, leading to a restricted (R) version of $A_Q$ given by
            \bn A_{Q(R)}\hskip-1.4em&&=\tint_0^T\<p(t),q(t)|\s[\s i\s\hbar\d/\d t-\H\s]\s|p(t),q(t)\>\,dt\no\\
               &&=\tint_0^T[\s p(t)\s {\dot q}(t)-H(p(t),q(t))\s]\,dt\;, \label{AQR}\en
               in which $H(p,q)\equiv \<p,q|\s\H\s|p,q\>$ and $i\hbar\<p,q|\s\s d|p,q\>=\<0|[(P+p\one)\s dq-Q\s dp\s]\s|0\>=p\s\s dq$. The restricted quantum action has the appearance of $A_C$, and, besides a new parameter ($\hbar$), results in Hamiltonian classical mechanics apart from one issue. In standard classical mechanics, $p$ and $q$ represent {\it sharp} values, while in (\ref{AQR}) the values of $p$ and $q$ are {\it mean} values since $p=\<p,q|\s P\s|p,q\>$ and $q=\<p,q|\s Q\s|p,q\>$, each variable having a standard deviation of $\sqrt{\hbar/2}$. However, we may question the assumption of sharp values for $p$ and $q$ since no one has measured their values to a precision (say) of $10^{137}$.   
           Thus we conclude that the restricted quantum action functional {\it IS} the classical action functional,
           and, since $\hbar>0$ still, we can claim that {\it the classical theory} is a subset of  {\it the quantum theory},
           meaning that {\it the classical and quantum theories coexist!} The equations of motion that follow from (\ref{AQR}) will generally include the parameter $\hbar$. If we are interested in the idealized classical ($c$) Hamiltonian for which $\hbar=0$, we may subsequently consider
              \bn H_c(p,q)\equiv \lim_{\hbar\ra0} H(p,q)=\lim_{\hbar\ra0}\<p,q|\s\H\s|p,q\>\;,\en
           but it is likely to be more interesting to retain $H(p,q)$ with $\hbar>0$ and see what difference $\hbar$ makes, if any, in the usual classical story.

           Changing coordinates by canonical coordinate transformations is sometimes problematic in traditional quantum theory. However, no such problems arise in the present formulation.
           Canonical coordinate transformations
           involve the change  $(p,q)\ra(\tp,\tq)$ for which $\{\tq,\tp\s\}=1$ as  well as
           $p\s\s dq=\tp\s\s d\tq+d{\widetilde G}(\tp,\tq)$, where ${\widetilde G}$ is called the generator of the canonical transformation. The coherent states $|p,q\>$ were chosen as a map from a particular point in phase space labeled by $(p,q)$ to a unit vector in Hilbert space. We want the same  result in the new coordinates, so we define a canonical coordinate transformation of the coherent states to be as a scalar, namely
             \bn |\tp,\tq\>\equiv |p(\tp,\tq),q(\tp,\tq)\>=|p,q\>\;, \en
             and thus the restricted quantum action functional becomes
             \bn A_{Q(R)}\hskip-1.4em&&=\tint_0^T\<\tp(t),\tq(t)|\s[\s i\hbar\s \d/\d t-\H\s]\s|\tp(t),\tq(t)\>\,dt\no\\
               &&=\tint_0^T\s[\s \tp(t)\s {\dot\tq}(t)+{\dot{\widetilde G}}(\tp(t),\tq(t))-{\widetilde H}(\tp(t),\tq(t))\s]\,dt\;,\en
               where ${\widetilde H}(\tp,\tq)\equiv H(p,q)$. Evidently this action functional leads to the proper form of Hamilton's equations after a canonical transformation. Observe that {\it no} change of the quantum operators occurs from these canonical transformations.

                The original coordinates we have chosen, i.e., $(p,q)$, enter the unitary operators $U(q)$ and $V(p)$ in (\ref{ccs}) as so-called canonical group coordinates \cite{GROUP}. In these coordinates, and assuming that $P$ and $Q$ are irreducible, it follows that
                  \bn H(p,q)\hskip-1.4em&&=\<p,q|\s \H(P,Q)\s|p,q\>\no\\
                  &&=\<0|\s \H(P+p\one,Q+q\one)\s|0\>\no\\
                  &&=\H(p,q)+{\cal O}(\hbar;p,q)\;. \label{AAA} \en
                  This expression confirms that in these coordinates $H(p,q)\ra\H=H(P,Q)$ up to terms ${\cal O}(\hbar)$. As is well known \cite{dirac}, in traditional quantum theory,  the special property represented by (\ref{AAA}) requires ``Cartesian coordinates''. Although that requirement is sometimes difficult to ascertain in canonical quantization, that is not the case in the proposed formalism. The Fubini-Study metric (which vanishes for vector variations that differ only in phase) for the canonical coherent states is 
                      \bn d\sigma^2(p,q)\hskip-1.4em&&\equiv (2\s\hbar)\,[\s\|\,d|p,q\>\|^2-|\<p,q|\s\s d|p,q\>|^2\s]\no\\
                                  &&= dp^2+dq^2\;,\en
                 which describes a flat, two-dimensional phase space in Cartesian coordinates. In different  coordinates, e.g.,  $(\tp,\tq)$, the metric will generally no longer be
                 expressed in Cartesian coordinates. Note carefully that this metric does not originate with the classical phase space but it describes the geometry of the sheet of coherent states in the Hilbert space of unit vectors. Of course, one can append such a metric to the classical phase space if one is so inclined.\footnote{One may imagine a future time when lowest-order effects in $\hbar$ may become observable by macroscopic measurements and rather than coherent states another set of, say, 
                 extended coherent states such as
                 \b  |p,q,a,b\>\equiv e^{\t -ia(P^2+Q^2)/\hbar}\,e^{\t -ib(PQ+QP)/\hbar}\,
                 e^{\t-iqP/\hbar}\,e^{\t ipQ/\hbar}\,|0\>\e
                 could be considered relevant. The use of these states in a restricted quantum action
                 would lead to a noncanonical set of extended equations of motion that generally would 
                 provide a better description of quantum evolution than is available by the coherent states alone. Before that time comes, however, we can treat such states as approximate quantum states without any mention of their possible classical significance.}

    \subsection*{Enhanced Affine Quantization}
             We now discuss another variation on conventional quantization methods offered by the general
            principles of enhanced quantization. Starting with the usual classical theory, we note that multiplying the Poisson bracket $1=\{q,p\}$ by $q$ leads to $q=q\{q,p\}=\{q,pq\}$,
            or $\{q,d\}=q$ where $d\equiv pq$. The two variables $d$ and $q$ form a Lie algebra and are worthy of consideration as a new pair of classical variables even though they are not canonical coordinates. It is also possible to restrict $q$ to $q>0$ or $q<0$ consistent with $d$, a variable that acts to {\it dilate} $q$ and not {\it translate} $q$ as $p$ does. Let us develop the quantum story featuring these variables.

            Take the Heisenberg commutation relation $i\hbar\one=[Q,P]$ and multiply both sides by $Q$ leading to $i\hbar\s Q=Q\s[Q,P]=[Q,QP]=[Q,\half(PQ+QP)]$, or stated otherwise,
              \bn [Q,D]=i\hbar Q\;,\hskip4em D\equiv \half(PQ+QP)\;. \en
              This is called the affine commutation relation and $D$ and $Q$ are called affine variables.
                Again, $D$ acts to {dilate} $Q$ while $P$ acts to {translate} $Q$, just like the classical variables. Observe that the affine commutation relation has been {\it derived} from the canonical commutation relation. Even though the original $P$ and $Q$ were irreducible,
              the operators $D$ and $Q$ are reducible. Indeed, there are three inequivalent irreducible representations: one with $Q>0$, one with $Q<0$ and one with $Q=0$, and all three involve representations that are self adjoint \cite{GELF}. The first two irreducible choices are the most interesting, and,  
               for the present, we focus on the choice $Q>0$. Conventional quantization techniques would suggest we start with the classical variables $d$ and $q$ and promote them to operators $D$ and $Q$ that fulfill the affine commutation relation. Although possible, that approach would seem to have little to do with canonical theories, either classical or quantum.
              However, if we adopt an enhanced quantization approach, that is no longer the case.

              Clearly the dimensions of $D$ are those of $\hbar$, and for convenience we treat $Q$ as dimensionless (or replace $Q$ by $Q/q_0$, with $q_0>0$, and use units where $q_0=1$).
              To build the affine coherent states based on the affine variables, we first must choose
              a fiducial vector. We choose the unit vector $|\tb\>$ as an extremal weight vector which is a solution of the equation
              \bn  [\s (Q-1)+(i/\tb)D\s]\s|\tb\>=0\;,\en
             where $\tb$ has the units of $\hbar$, and it follows that $\<\tb|Q|\tb\>=1$ and $\<\tb|D|\tb\>=0$. The affine coherent states are now defined as
               \bn |p,q\>\equiv e^{\t ip\s Q/\hbar}\,e^{\t -i\ln(q)\s D/\hbar}\,|\tb\>\;,\en
               for $(p,q)\in{\mathbb R}\times{\mathbb R}^+$, i.e., $q>0$. Despite a similar notation, all coherent states in this section are affine coherent states; indeed, the limited range of $q$ is an implicit distinction between the affine and canonical coherent states.

               It is noteworthy that the geometry of the affine coherent states is different from the geometry of the canonical coherent states. To see this we consider the Fubini-Study metric  that holds for the affine coherent states, namely 
                \bn d\sigma^2(p,q)\hskip-1.4em&&\equiv (2\s\hbar)\s[\s\|\s\s d|p,q\>\|^2-|\<p,q|\s\s d|p,q\>|^2\s]\no\\
                   &&=\tb^{-1}\s q^2\,dp^2+\tb\s q^{-2}\,dq^2\;. \en
                    This metric describes a two-dimensional space of constant negative curvature: $-2/\tb$.  Since $q>0$, it follows that this (Poincar\'e) half plane  is geodesically complete. Just as in the canonical case, however,
                   this metric describes the geometry of the sheet of coherent states, and does not  originate from the classical phase space.

               We are now in position to examine the enhanced quantization of such a system.  Although the basic affine variables are $D$ and $Q$, it is sometimes convenient to use $P$ as well, even
               though it cannot be made self adjoint. Thus, the Hamiltonian operators $\H$ for such variables may be taken as $\H'(D,Q)$ or as $\H(P,Q)\equiv \H'(D,Q)$.  We start with
               the quantum action functional for affine variables given by
                 \bn A'_Q=\tint_0^T\<\psi(t)|\s[\s i\s\hbar\s \d/\d t-\H\s]\s|\psi(t)\>\,dt\;, \en
                 which leads to the same formal Schr\"odinger's equation under stationary variation.
                 But what classical system does this correspond to? As part of enhanced quantization, that answer arises when we restrict the domain of the quantum action functional to the affine coherent states,
                 \bn A'_{Q(R)}\hskip-1.4em&&=\tint_0^T\<p(t),q(t)|\s[\s i\s\hbar\s \d/\d t-\H\s]\s|p(t),q(t)\>\,dt \no\\
                 &&=\tint_0^T[\s -q(t)\s\s {\dot p}(t)-H(p(t),q(t))\s]\,dt\;,\en
              where  $\<p,q|\s\s d|p,q\>=\<\tb|\s(\s -q\s Q\s\s dp+ d\s\ln(q)\s D)|\tb\>=-q\s\s dp$ and $H(p,q)\equiv \<p,q|\H|p,q\>$. In short, {\it the classical limit of enhanced affine quantization is a canonical theory!} Thus, if we deal with a classical system for which the physics requires that $q>0$, then enhanced quantization, which requires self adjoint generators,  cannot proceed with the canonical variables $P$ and $Q$ (because $P$ is not self adjoint), but it must use the self-adjoint affine variables $D$ and $Q$. To emphasize the canonical nature of the enhanced classical theory, we note that we can again introduce canonical coordinate transformations such that $|\tp,\tq\>\equiv|p(\tp,\tq),q(\tp,\tq)\>=|p,q\>$,
              leading as before to an expression for $A_{Q(R)}$ involving the new coordinates $(\tp,\tq)$,
              but without any change of the quantum operators whatsoever.

              To complete the story we need to give further details regarding the dynamics. In particular, we note that
                \bn  H(p,q)=\<\tb|\s\H'(D+p\s q\s Q,q\s Q)\s|\tb\>=\<\tb|\s \H(P/q+p\one,q\s Q)\s|\tb\>\;.\en
              In order to proceed further we let $|x\>$, $x>0$, where $Q\s|x\>=x\s|x\>$,  be a basis in which $Q\ra x$ and $D\ra -i\s\hbar(x\s\d/\d x+\half)$. With $M$ a normalization factor, it follows that
              \bn \<x|\tb\>=M\,x^{\tb/\hbar-1/2}\,e^{-(\tb/\hbar)\s x}\;,\en
              for which $\<p,q|\s Q^n\s|p,q\>=\<\tb|\s (q\s Q)^n\s|\tb\>=q^n+{\cal O}(\hbar)$ and
              $\<p,q|\s P^2\s|p,q\>=\<\tb|\s (P/q+p\one)^2\s|\tb\>=p^2+C_2/q^2$, where $C_2\equiv\<\tb|\s P^2\s|\tb\>\propto \hbar^2$ and $\<\tb|\s P\s|\tb\>=0$.

              Thus, as an example, the classical, one-dimensional ``Hydrogen atom'' with  $H_c=p^2/(2m)-e^2/q$, $q>0$, actually leads to singularities since, generally, $q(t)\ra0$ in a finite time. On the basis of an affine quantization, however, the enhanced
              classical Hamiltonian $H(p,q)=p^2/(2m)-C_1/q+C_2/(2m\s q^2)$, 
              where $C_1=e^2+{\cal O}(\hbar)$, but, more importantly, $C_2/(2m)\simeq (\hbar^2/m\s e^2)\,C_1$, and the coefficient is recognized as the Bohr radius! Thus, for this model, all conventional classical singularities are removed by quantum corrections to the classical theory using $H(p,q)$. However, if we first let $\hbar\ra0$, leading to $H_c(p,q)$, the singularities will return.

              It is noteworthy that simple models of cosmic singularities of the gravitational field, which exhibit classical singularities, have recently been shown to have classical bounces instead, based on enhanced affine quantization \cite{zon}.

 \subsection*{Enhanced Spin Quantization}
              For completeness we ask whether there is still another pair of quantum operators that can also serve as partners for an enhanced quantization of classical phase space. Indeed, there is another family, and we briefly outline its properties. Consider a set of three spin operators which
satisfy $[S_1,S_2]=i\hbar\s S_3$ plus cyclic permutations. We choose an irreducible representation where $\Sigma_{j=1}^3 S_j^2=\hbar^2\s s(s+1)$, the spin $s\in\{1/2,1,3/2,2,\ldots\}$, and the representation space is a (2\s s+1)-dimensional Hilbert space. Normalized 
eigenvectors and their eigenvalues for $S_3$ are given by $S_3\,|s,m\>=m\s\hbar\,|s,m\>$  
for $-s\le m\le s$. We choose $|s,s\>$ as our
fiducial vector, which is also an extremal weight vector since $(S_1+i\s S_2)\s|s,s\>=0$. As spin coherent states we choose
\bn |\th,\p\>\equiv e^{\t -i\s\p\s S_3/\hbar}\,e^{\t-i\th\s S_2/\hbar}\,|s,s\>\;,\en  
for $0\le\th\le\pi$ and $-\pi<\p\le\pi$;
they may also be described by
    \bn |p,q\>\equiv e^{\t-i\s (q/(s\hbar)^{1/2})\s S_3/\hbar}\,e^{\t-i\s\cos^{-1}(p/(s\hbar)^{1/2})\s S_2/\hbar}\,|s,s\>\;,\en 
    where $-(s\hbar)^{1/2}\le p\le (s\hbar)^{1/2}$ and $-\pi\s(s\hbar)^{1/2}< q\le \pi\s(s\hbar)^{1/2}$.
    The quantum action functional is
    \bn A''_Q=\tint_0^T \<\psi(t)|\s[\s i\hbar\d/\d t-\H\s]\s|\psi(t)\>\,dt\;,\en
    and with $|\psi(t)\>\ra|\th(t),\p(t)\>$, it becomes
    \bn A''_{Q(R)}\hskip-1.4em&&=\tint_0^T[\s s\hbar\s \cos(\th)\s\s {\dot\p}(t)-H(\th(t),\p(t))\s]\,dt\no\\
                  &&=\tint_0^T[\s p(t)\s\s{\dot q}(t)-H(p(t),q(t))\s]\,dt\;. \en
    Canonical transformations follow the same pattern as before. The Fubini-Study metric on the spin coherent state sheet in the $(2s+1)$-dimensional Hilbert space is given by 
      \bn d\sigma^2\hskip-1,4em&&\equiv (2\s\hbar)\s[\s \|\s\s d|\th,\p\>\|^2-|\<\th,\p|\s\s d|\th,\p\>|^2\s]\no\\
        &&=(s\hbar)\s [\s d\th^2+\sin(\th)^2\s d\p^2\s]\no\\
        &&= [1-p^2/(s\hbar)]^{-1}\s dp^2+[1-p^2/(s\hbar)]\s dq^2\;,\en
        corresponding to a two-dimensional sphere of radius $(s\hbar)^{1/2}$. If we let $\hbar\ra0$, the whole story in this section vanishes.

 \subsection*{Conclusion}
              The foregoing discussion of one-dimensional mechanical examples shows how enhanced quantization techniques offer a new outlook on the very process of quantization itself. The principal motivating force behind enhanced quantization is to develop a formalism that lets both the classical and quantum theories coexist. This requires decoupling the usual assignment of classical and quantum variables by the promotion of one into the other. Instead, using enhanced quantization, in which the classical and quantum systems do coexist, we have seen the added benefit that the traditional difficulties surrounding canonical coordinate transformations disappear. Although we have focused on an enhanced affine quantization for which $q>0$, and thus $Q>0$---and even that domain can be readily changed to $q>-\gamma$ and $Q>-\gamma\one$---it is also possible to study affine models for which both $q$ and the spectrum of $Q$ run over the entire real line, which then lets enhanced affine quantization compete with canonical quantization itself.

            The  extension of enhanced quantization to multiple degrees of freedom is straightforward, and  enhanced quantization may also be used for field theories. Noteworthy in this latter category are several examples:
               Ultralocal and covariant scalar field theories, which lead to triviality (and nonrenormalizability) by traditional quantization methods, have been shown to become nontrivial and well defined when quantized by enhanced affine methods \cite{acs-IOP,IOP,MOSCOW}.
              Additionally, Einstein gravity features variables such as the spatial metric that satisfy certain positivity conditions, which suggests that they should be quantized by enhanced affine quantization. Such studies have already begun; see\cite{AQG3} and references therein.


\begin{thebibliography}{99}

 \bibitem{CRT2} J.R.~Klauder, ``Continuous-Representation Theory II.  Generalized
        Relation Between
   Quantum and Classical Dynamics", J. Math. Phys. {\bf 4}, 1058-1073 (1963).

  \bibitem{CRT3}   J.R.~Klauder, ``Continuous-Representation Theory III.  On Functional
       Quantization of Classical Systems", J. Math. Phys. {\bf 5}, 177-187 (1964).

   \bibitem{KLASK}   J.R.~Klauder and E.W.~Aslaksen, ``Elementary Model for Quantum
        Gravity", Phys. Rev. D {\bf 2}, 272-276 (1970).

  \bibitem{GROUP}  P.M.~Cohn, ``Lie Groups'', (Cambridge University Press, London, 1961).

  \bibitem{dirac}  P.A.M.~Dirac, ``The Principles of Quantum Mechanics'', (Clarendon Press,  Oxford, 1947), page 114.

  \bibitem{GELF} I.M.~Gel'fand and M.I.~Naimark, Dokl. Akad. Nauk. SSSR 55, 570 (1947); E.W.~Aslaksen and J.R.~Klauder, J. Math. Phys. {\bf 9}, 206 (1968).

  \bibitem{zon} M. Fanuel and S. Zonetti, ``Affine Quantization and the Initial Cosmological Singularity'',
     arXiv:1203.4936.

  \bibitem{acs-IOP} J.R.~Klauder, ``The Utility of Affine Variables and Affine Coherent States'', arXiv:1108.3380.



  \bibitem{IOP} J.R.~Klauder, ``Scalar Field Quantization Without Divergences In All Spacetime Dimensions'',
  J.~Phys.~A: Math.~Theor.~{\bf 44}, 273001 (30pages) (2011).

      \bibitem{MOSCOW} J.R.~Klauder, ``Enhanced Quantum Procedures that Resolve Difficult Problems'', in
      preparation.

     \bibitem{AQG3}J.R.~Klauder,  ``Recent Results Regarding Affine Quantum Gravity'', arXiv:1203.0691.


\end{thebibliography}
\end{document}